\documentclass[aps,prb,twocolumn,showpacs]{revtex4}
\usepackage{graphicx}
\usepackage{bm}
\begin{document}
\title{Local phonon density of states in an elastic substrate}
\author{Michael R. Geller}
\affiliation{Department of Physics and Astronomy, University of Georgia, Athens, Georgia 30602-2451}

\date{June 3, 2004}

\begin{abstract}
The local, eigenfunction-weighted acoustic phonon density of states (DOS) tensor is calculated for a model substrate consisting of a semi-infinite isotropic elastic continuum with a stress-free surface. On the surface, the local DOS is proportional to the square of the frequency $\omega$, as for the three-dimensional Debye model, but with a constant of proportionality that is considerably enhanced compared to the Debye value, a consequence of the Rayleigh surface modes. The local DOS tensor at the surface is also anisotropic, as expected. Inside the substrate the local DOS is both spatially anisotropic and non-quadratic in frequency. However, at depths large compared with $v_{\rm l}/\omega$, where $v_{\rm l}$ is the bulk longitudinal sound velocity, the local DOS approaches the isotropic Debye value. The results are applied to a Si substrate.
\end{abstract}

\maketitle

\section{introduction and elastic substrate model\label{introduction section}}

A variety of surface and near-surface thermodynamic and lattice dynamical phenomena are governed by the phonon density of states (DOS), the number of vibrational modes per unit frequency, per unit volume of solid. In an inhomogeneous system, however, it is possible to distinguish between the ``global'' DOS, which makes reference to only to the vibrational-mode frequencies of the entire system, and a local, eigenfunction-weighted DOS. In systems with a small {\it mechanical} surface-to-volume ratio, the volume including all parts of the system and environment participating in the vibrations, the global DOS is essentially unchanged from that of a bulk system. The eigenfunction-weighted DOS is interesting both because it is relevant to a variety of physical phenomena, and because it contains local, position-dependent mechanical and vibrational information about the system.

In this paper, we calculate the local phonon DOS tensor for a model substrate consisting of a semi-infinite isotropic elastic continuum with a stress-free surface, appropriate for low-frequency acoustic phonons in real solids with an exposed clean surface. The elastic continuum is also assumed to be linear and nonpolar. The free surface is taken to lie in the $xy$ plane, with the substrate occupying the $z>0$ half-space. Periodic boundary conditions are applied in the lateral directions, on an area ${\sf A}$. Normalized vibrational eigenfunctions have been obtained for this geometry in a classic paper by Ezawa.\cite{Ezawa}

\section{local phonon DOS tensor \label{definition section}}

We begin by defining the local DOS tensor $g_{ij}({\bf r},\omega)$, the quantity to be calculated in this paper. The Lagrangian density for a linear, nonpolar, isotropic elastic continuum is
\begin{equation}
{\cal L} = {\textstyle{1 \over 2}} \rho \, (\partial_t {\bf u})^2 - {\textstyle{1 \over 2}} \lambda \, u_{ii}^2 - \mu \, u_{ij}^2,
\label{lagrangian} 
\end{equation}
where ${\bf u}({\bf r},t)$ is the displacement field, and where $u_{ij} \equiv (\partial_i u_j + \partial_j u_i)/2$ is the strain tensor. $\rho$ is the mass density, and $\lambda$ and $\mu$ are the standard Lam\'e coefficients, related to the bulk longitudinal and transverse sound velocities $v_{\rm l}$ and $v_{\rm t}$ according to $v_{\rm l} = \sqrt{ (\lambda + 2 \mu) / \rho}$ and $v_{\rm t} \equiv \sqrt{\mu / \rho}.$ The equation of motion following from Eq.~(\ref{lagrangian}) is
\begin{equation}
\partial_t^2 {\bf u} - v_{\rm l}^2 {\bm \nabla} ({\bm \nabla} \cdot {\bf u}) +  v_{\rm t}^2 {\bm \nabla} \times {\bm \nabla} \times {\bf u} = 0.
\label{elasticity equation}
\end{equation}

It will also be useful to define an elastic stress tensor $T_{ij}$ from the continuity equation $\partial_t \Pi_i + \partial_j T_{ij} = 0$ for momentum density ${\bf \Pi} \equiv \rho \, \partial_t {\bf u}$. In an isotropic elastic medium, it follows from Eq.~(\ref{elasticity equation}) that 
\begin{equation}
T_{ij} = - \lambda ({\bm \nabla} \cdot {\bf u}) \, \delta_{ij} - 2 \mu \, u_{ij}  = - \, c_{ijkl} \, u_{kl},
\label{stress tensor}
\end{equation}
where 
\begin{equation}
c_{ijkl} = \lambda \, \delta_{ij} \delta_{kl} + \mu (\delta_{ik} \delta_{jl} + \delta_{il} \delta_{jk}) 
\label{elastic tensor}
\end{equation} 
is the corresponding elastic tensor. The boundary condition at the $z=0$ exposed planar surface ${\sf S}$ is
\begin{equation}
T_{iz}({\bf r}) = 0, \ \ \ {\bf r} \in {\sf S}, \ \ \  i=x,y,z.
\end{equation}

To quantize the elastic waves we require the displacement field and momentum density to satisfy the canonical commutation relation
\begin{equation}
[ u_i({\bf r}) , \Pi_j({\bf r}') ] = i \hbar \delta_{ij}  \delta({\bf r}-{\bf r}').
\label{commutation relation}
\end{equation}
Therefore, we can expand the displacement field in a basis of bosonic creation and annihilation operators according to
\begin{equation}
{\bf u}({\bf r}) = \sum_n \sqrt{\hbar \over 2 \rho \omega_n} \ \bigg[  a_n \, {\bf f}_n({\bf r}) +  a_n^\dagger \,  {\bf f}^*_n({\bf r})  \bigg],
\label{displacement field expansion}
\end{equation}
where the ${\bf f}_n({\bf r})$ are vibrational eigenfunctions, defined to be time-periodic solutions of the elasticity equation (\ref{elasticity equation}), in the presence of the stated boundary conditions, and normalized according to $\int \!  d^3r \ {\bf f}_n^* \cdot {\bf f}_{n'} = \delta_{nn'}.$ Here the integration is over the volume ${\sf V}$ of the system. The vibrational eigenfunctions are assumed to be complete, so that $\sum_n f^i_n({\bf r}) \, {f_n^j}^* \! ({\bf r}') = \delta_{ij} \, \delta({\bf r - r'}).$

The quantity we shall calculate in this paper is\cite{definition footnote}
\begin{equation}
g_{ij}({\bf r},\omega) \equiv \sum_n \delta(\omega - \omega_n ) \, f_n^i({\bf r}) \, [f_n^j({\bf r})]^* \! ,
\label{g definition}
\end{equation}
where $i$ and $j$ are Cartesian tensor indices. This quantity weights each vibrational mode $n$ by the square of the amplitude of the eigenfunction at position ${\bf r}$. The trace of $g_{ij}({\bf r},\omega)$ also contains physically interesting information, namely
\begin{equation}
{\rm Tr} \, g_{ij}({\bf r},\omega) = \sum_i g_{ii}({\bf r},\omega)  = \sum_n \delta(\omega - \omega_n ) \, |{\bf f}_n({\bf r})|^2 \! .
\label{trace expression}
\end{equation}
Now, averaging the trace of $g_{ij}({\bf r},\omega)$ over the system volume ${\sf V}$ leads to
\begin{equation}
{1 \over {\sf V}} \int d^3r \,  {\rm Tr} \, g_{ij}({\bf r},\omega) = {1 \over {\sf V}} \sum_n \delta(\omega - \omega_n ),
\label{average trace expression}
\end{equation}
which is the ordinary intensive global DOS, i.e., the number of vibrational modes per unit frequency per volume.

In an isotropic system with translational invariance, the local DOS is independent of position, and Eq.~(\ref{g definition}) simplifies to
\begin{equation}
g_{ij}({\bf r},\omega) = {\delta_{ij} \over 3{\sf V}} \sum_n \delta(\omega - \omega_n ), 
\label{translationally invariant g}
\end{equation}
where a three-dimensional system is assumed. Because there is no position dependence here, the trace of this quantity yields the global DOS [the right-hand-side of Eq.~(\ref{average trace expression})].

Is the local DOS defined Eq.~(\ref{g definition}) a quantum mechanical quantity? One the one hand, it does not involve $\hbar$, and in that sense it is not quantum mechanical. On the other hand, however, the normalization assumed for the ${\bf f}_n({\bf r})$ is of quantum-mechanical origin: Assuming the conventional bosonic commutation relations for the $a_n$ and $a_n^\dagger$ appearing in the expansion of Eq.~(\ref{displacement field expansion}), the normalization condition on the vibrational eigenfunctions is required for Eq.~(\ref{commutation relation}) to be satisfied.\cite{normalization footnote} Therefore, the local DOS defined Eq.~(\ref{g definition}) naturally shows up in quantum mechanical calculations of the vibrational properties of inhomogeneous systems.

\section{local dos in the three-dimensional bulk limit \label{bulk limit section}}

In what follows it will be useful to recall the well-known Debye DOS formula. The spectrum and normalized vibrational eigenfunctions for a three-dimensional isotropic elastic continuum, with periodic boundary conditions applied to the surface of a three-dimensional volume ${\sf V}$, are
\begin{equation}
\omega_{m{\bf k}} = v_m \, |{\bf k}| \ \ \ \ \ {\rm and} \ \ \ \ \ {\bf f}_{m{\bf k}}({\bf r}) =  {e^{i {\bf k} \cdot {\bf r}} \over \sqrt{\sf V}} 
\, {\bm \epsilon}_{m{\bf k}},
\end{equation}
where $m \! = \! ({\rm l}, {\rm t}_1, {\rm t}_2)$ labels the longitudinal and two transverse branches, the sound velocities are the bulk values quoted in Sec.~\ref{definition section}, and the ${\bm \epsilon}_{m{\bf k}}$ are unit polarization vectors forming an orthogonal basis. ${\bm \epsilon}_{{\rm l}{\bf k}}$ is orietnted in the ${\bf k}$ direction, and ${\bm \epsilon}_{{\rm t}_1{\bf k}}$ and ${\bm \epsilon}_{{\rm t}_2{\bf k}}$ are perpendicular to ${\bm \epsilon}_{{\rm l}{\bf k}}$ and to each other.

In this case, Eq.~(\ref{g definition}) yields
\begin{equation}
g_{ij}({\bf r},\omega) = {1 \over {\sf V}} \sum_{m{\bf k}} \delta(\omega - \omega_{m{\bf k}}) \, \epsilon^i_{m{\bf k}} \epsilon^j_{m{\bf k}}. 
\end{equation}
The fact that $\omega_{m{\bf k}}$ is independent of the direction of ${\bf k}$ allows the product of polarization vectors to be averaged over the unit sphere, leading to $\delta_{ij}/D$ in $D$ dimensions. We then obtain
\begin{equation}
g_{ij}({\bf r},\omega) = {\delta_{ij} \over 3{\sf V}} \sum_{m{\bf k}} \delta(\omega - \omega_{m{\bf k}}), 
\end{equation}
an example of the form given in Eq.~(\ref{translationally invariant g}). Therefore, in the ${\sf V} \rightarrow \infty$ limit, 
\begin{equation}
g_{ij}({\bf r},\omega) = {\delta_{ij} \over 3} \times {\omega^2 \over 2 \pi^2} \bigg({1 \over v^3_{\rm l}} + {2 \over v^3_{\rm t}} \bigg). 
\label{bulk g}
\end{equation}

Eq.~(\ref{bulk g}) is the conventional three-dimensional Debye DOS tensor for acoustic phonons. Of course, in a real solid the continuum approximation breaks down at short wavelengths and the spectrum is cut off at high frequencies.

\section{local DOS in semi-infinite elastic substrate with a stress-free surface \label{substrate section}}

The evaluation of the local DOS for our substrate model requires the vibrational eigenfunctions of a semi-infinite isotropic elastic continuum with a stress-free surface, which have been obtained by Ezawa.\cite{Ezawa} The modes are labeled by a branch index $m$, taking the five values \ SH, $+$, $-$, 0, and R, by a two-dimensional wavevector ${\bf K}$ in the plane defined by the surface, and by a parameter $c$ with the dimensions of velocity that is continuous for all branches except the Rayleigh branch $m = {\rm R}$. The range of the parameter $c$ depends on the branch $m$, and is given in Table \ref{c table}. The frequency of mode $(m, {\bf K}, c)$ is $\omega_{m {\bf K} c} = c K,$ where $K \equiv |{\bf K}|$ is the magnitude of the two-dimensional wavevector. As stated above, we shall assume periodic boundary conditions in the $x$ and $y$ direction applied to a finite area ${\sf A}$.\cite{area footnote}

\begin{table}[!h]
{\centering
\begin{tabular}{|c|c|}  \hline
$m$  & \ \ \ \ \ \ range of $c$ \ \ \ \ \   \\ \hline
SH    & $[v_{\rm t} , \infty]$  \\
$\pm$ & $[v_{\rm l} , \infty]$  \\
0     & $[v_{\rm t} , v_{\rm l}]$  \\
R     & \, $c_{\rm R}$ \, (discrete)  \\  \hline
\end{tabular}
\par}
\caption{Values of the parameter $c$ for the five branches of vibrational modes of a semi-infinite substrate. The Rayleigh branch $m={\rm R}$ has a single value of $c$, which we denote by $c_{\rm R}$.}
\label{c table}
\end{table}

\begin{widetext}

The definition of the local DOS given in Eq.~(\ref{g definition}) applies to a system with a discrete spectrum. The generalization of Eq.~(\ref{g definition}) to our model, which has a mixed discrete-continuous spectrum, is
\begin{equation}
g_{ij}({\bf r},\omega) =  \sum_{\bf K} \delta(\omega - c_{\rm R} K ) \, f_{{\rm R}{\bf K}}^i({\bf r}) \, [f_{{\rm R}{\bf K}}^j({\bf r})]^* 
+ \sum_{\bf K} \sum_{m \neq {\rm R}} \int_{\Gamma_{\! m}} \! \! \! dc \ \delta(\omega - cK) \, f_{m{\bf K}c}^i({\bf r}) \, 
[f_{m{\bf K}c}^j({\bf r})]^* \! .
\label{substrate g definition}
\end{equation}
The first term accounts for the Rayleigh surface wave branch $m \! = \! {\rm R}$, and these vibrational mode eigenfunctions are normalized as in Sec.~\ref{definition section}. The value of $c$ for the Rayleigh branch, $c_{\rm R},$ is calculated below. The second term in Eq.~(\ref{substrate g definition}) accounts for the four remaining branches $m  = {\rm SH}, + , -,$ and $0,$ having continuous values of $c$. Their eigenfunctions are instead normalized according to $\int \!  d^3r \ {\bf f}_{m {\bf K} c}^* \cdot {\bf f}_{m' {\bf K}' c'} = \delta_{mm'} \, \delta_{\bf K K'} \, \delta(c - c').$ Furthermore, $\Gamma_{\! m}$ is the integration domain given in Table \ref{c table}.

\end{widetext}

We turn now to an evaluation of Eq.~(\ref{substrate g definition}). The contribution from each branch in Table \ref{c table} will be discussed separately in the sections to follow.

\subsection{SH branch}

First we consider the $m={\rm SH}$ branch, for which\cite{Ezawa}
\begin{equation}
{\bf f}_{\rm SH} = \sqrt{2cK \over \pi v^2_{\rm t} \beta {\sf A}} \ \cos(\beta K z) \ {\bf e}_z \! \times \! {\bf e}_{\rm K}  \ e^{i {\bf K} \cdot {\bf r}},
\end{equation}
where
\begin{equation}
\beta(c) \equiv \sqrt{(c/v_{\rm t})^2 - 1},
\end{equation}
and where ${\bf e}_{\rm K} \equiv {\bf K}/K$ is a unit vector in the ${\bf K}$ direction. This mode is polarized in the $xy$ plane, perpendicular to ${\bf K}$. The contribution made by this branch to the local DOS is
\begin{eqnarray}
g_{ij}^{({\rm SH})}({\bf r},\omega) &=& {2 \omega \over \pi v_{\rm t}^2 {\sf A}} \sum_{\bf K} \int_{{\rm v}_t}^\infty \! \! \! dc \ {\cos^2(\omega \beta z/c) \over \beta} \nonumber \\
&\times& \delta(\omega - cK) \ ({\bf e}_z \! \times \! {\bf e}_{\rm K})^i ({\bf e}_z \! \times \! {\bf e}_{\rm K})^j.
\end{eqnarray}
Taking the ${\sf A} \rightarrow \infty$ limit, and making use of the {\it two}-dimensional DOS formula
\begin{equation}
{1 \over {\sf A}} \sum_{\bf K} \delta(\omega - cK) = \int \! {d^2 K \over 4 \pi^2} \, \delta(\omega - cK) = {\omega \over 2 \pi c^2},
\label{2D DOS}
\end{equation}
leads to
\begin{equation}
g_{ij}^{({\rm SH})}({\bf r},\omega) = {\omega^2 \over \pi^2 v_{\rm t}^2} \int_{{\rm v}_t}^\infty \! \! \! dc \ {\cos^2(\omega \beta z/c) \over \beta c^2} \ 
{\overline{({\bf e}_z \! \times \! {\bf e}_{\rm K})^i ({\bf e}_z \! \times \! {\bf e}_{\rm K})^j}},
\end{equation}
where
\begin{equation}
{\overline{({\bf e}_z \! \times \! {\bf e}_{\rm K})^i ({\bf e}_z \! \times \! {\bf e}_{\rm K})^j}} \equiv \int \! {d \Omega \over 2 \pi} \ ({\bf e}_z \! \times \! {\bf e}_{\rm K})^i ({\bf e}_z \! \times \! {\bf e}_{\rm K})^j
\end{equation}
denotes a two-dimensional average over the directions of ${\bf K}$. Finally, using
\begin{equation}
{\overline{({\bf e}_z \! \times \! {\bf e}_{\rm K})^i ({\bf e}_z \! \times \! {\bf e}_{\rm K})^j}} = {\delta_{ij} \over 2} \big(1-\delta_{iz}\big)
\end{equation}
leads to
\begin{equation}
g_{ij}^{({\rm SH})}({\bf r},\omega) = {\omega^2 \over 2 \pi^2 v_{\rm t}^3} \delta_{ij} (1-\delta_{iz}) \, I_{\rm SH}\big({\textstyle{\omega z \over v_{\rm t}}}\big),
\end{equation}
where
\begin{equation}
I_{\rm SH}(Z) \equiv \int_1^\infty \! \! \! ds \ {\cos^2(Z\sqrt{s^2-1} / s) \over s^2 \sqrt{s^2-1}}. 
\end{equation}
Below we will make use of the large $Z$ limit of $I_{\rm SH}(Z)$, which is
\begin{equation}
\lim_{Z \rightarrow \infty} I_{\rm SH}(Z) = {1 \over 2}.
\end{equation}

\subsection{$\pm$ branches}

For the $m = + $ and $-$ branches the normalized vibrational eigenmodes 
are\cite{Ezawa} 
\begin{eqnarray}
{\bf f}_{\pm} &=&  \sqrt{K \over 4 \pi c {\sf A}} \bigg\{ \bigg[ \mp \alpha^{-{1 \over 2}} \, \big( e^{-i \alpha K z} - \zeta_{\pm} \, 
e^{i \alpha K z} \big)  \nonumber \\ 
&+&  i \beta^{1 \over 2} \big( e^{-i \beta K z} + \zeta_{\pm} \, e^{i \beta K z} \big) \bigg] {\bf e}_{\rm K} \nonumber \\
&+& \bigg[ \pm \alpha^{1 \over 2} \big( e^{-i \alpha K z} + \zeta_{\pm} \, e^{i \alpha K z} \big) \nonumber \\
&+& i \beta^{-{1 \over 2}} \big( e^{-i \beta K z} - \zeta_{\pm} \, e^{i \beta K z} \big) \bigg] {\bf e}_z \bigg\} e^{i {\bf K} \cdot {\bf r}},   
\label{pm modes}
\end{eqnarray}
where
\begin{equation}
\alpha(c) \equiv \sqrt{(c/v_{\rm l})^2 - 1} .
\end{equation}
Here $\zeta_{\pm}$ is a phase factor
\begin{equation}
\zeta_{\pm}(c) \equiv {[(\beta^2 - 1) \pm 2 i \sqrt{\alpha \beta} ]^2 \over (\beta^2 -1)^2 + 4 \alpha \beta},
\end{equation}
with $|\zeta_{\pm}(c)| = 1.$  We also note that
\begin{equation}
{\overline{e_{\rm K}^i e_{\rm K}^j}} = {\delta_{ij} \over 2} \big(1-\delta_{iz}\big).
\end{equation}

The $m = \pm$ branches therefore contribute an amount 
\begin{equation}
g_{ij}^{(+)}({\bf r},\omega) + g_{ij}^{(-)}({\bf r},\omega)
\end{equation}
to the local DOS, where
\begin{widetext}
\begin{equation}
g_{ij}^{(\pm)}({\bf r},\omega) = {\omega^2 \delta_{ij} \over 16 \pi^2 v_{\rm l}^3}  \bigg[  (1-\delta_{iz}) \, I_{\pm}\big({\textstyle{\omega z \over v_{\rm l}}}\big) 
+ 2 \, \delta_{iz}  J_{\pm}\big({\textstyle{\omega z \over v_{\rm l}}}\big) \bigg].
\end{equation}
Here
\begin{eqnarray}
I_{\pm}(Z) &\equiv& \int_1^\infty \! {ds \over s^4} \ \bigg| \big(s^2-1\big)^{-{1 \over 4}} \bigg[ e^{-i Z \sqrt{s^2-1} /s} - 
\bigg( { [{\textstyle{s^2 \over \nu^2}} - 2 \pm 2 i  (s^2-1)^{1 \over 4}({\textstyle{s^2 \over \nu^2}} - 2)^{1 \over 4} ]^2\over ({\textstyle{s^2 \over \nu^2}} - 2)^2 
+ 4 (s^2-1)^{1 \over 2}({\textstyle{s^2 \over \nu^2}} - 2)^{1 \over 2} } \bigg) 
e^{i Z\sqrt{s^2-1}/s} \bigg]  \nonumber \\
&\mp&  i \, \big({\textstyle{s^2 \over \nu^2}}-1\big)^{{1 \over 4}} \bigg[ e^{-i Z \sqrt{(s/\nu)^2-1} /s} + 
\bigg( { [{\textstyle{s^2 \over \nu^2}} - 2 \pm 2 i  (s^2-1)^{1 \over 4}({\textstyle{s^2 \over \nu^2}} - 2)^{1 \over 4}]^2 \over ({\textstyle{s^2 \over \nu^2}} - 2)^2 
+ 4 (s^2-1)^{1 \over 2}({\textstyle{s^2 \over \nu^2}} - 2)^{1 \over 2} } \bigg)
e^{i Z\sqrt{(s/\nu)^2-1}/s}\bigg] \bigg|^2 \! , 
\label{Ipm definition}
\end{eqnarray}
and
\begin{eqnarray}
J_{\pm}(Z) &\equiv& \int_1^\infty \! {ds \over s^4} \ \bigg| \big(s^2-1\big)^{{1 \over 4}} \bigg[ e^{-i Z \sqrt{s^2-1} /s} +
\bigg( {[ {\textstyle{s^2 \over \nu^2}} - 2 \pm 2 i  (s^2-1)^{1 \over 4}({\textstyle{s^2 \over \nu^2}} - 2)^{1 \over 4}]^2 \over ({\textstyle{s^2 \over \nu^2}} - 2)^2 
+ 4 (s^2-1)^{1 \over 2}({\textstyle{s^2 \over \nu^2}} - 2)^{1 \over 2} } \bigg) 
e^{i Z\sqrt{s^2-1}/s} \bigg]  \nonumber \\
&\pm&  i \, \big({\textstyle{s^2 \over \nu^2}}-1\big)^{-{1 \over 4}} \bigg[ e^{-i Z \sqrt{(s/\nu)^2-1} /s} - 
\bigg( { [{\textstyle{s^2 \over \nu^2}} - 2 \pm 2 i  (s^2-1)^{1 \over 4}({\textstyle{s^2 \over \nu^2}} - 2)^{1 \over 4}]^2 \over ({\textstyle{s^2 \over \nu^2}} - 2)^2 
+ 4 (s^2-1)^{1 \over 2}({\textstyle{s^2 \over \nu^2}} - 2)^{1 \over 2} } \bigg)
e^{i Z\sqrt{(s/\nu)^2-1}/s}\bigg] \bigg|^2 \! ,
\label{Jpm definition} 
\end{eqnarray}
\end{widetext}
where
\begin{equation}
\nu \equiv {v_{\rm t} \over v_{\rm l}}
\label{nu definition}
\end{equation}
is the ratio of transverse and longitudinal bulk sound velocities. It can be shown that
\begin{equation}
\lim_{Z \rightarrow \infty} I_{\pm}(Z) = {4 \over 3} + {2 \over 3 \nu^3} - {2 \over 3 \nu^3}\big(1-\nu^2\big)^{3/2}
\end{equation}
and
\begin{equation}
\lim_{Z \rightarrow \infty} J_{\pm}(Z) = {2 \over 3} +{4 \over 3 \nu^3}- {2 \over 3 \nu^3}\big(2+\nu^2\big)\sqrt{1-\nu^2}.
\end{equation}

\subsection{0 branch}

Next we consider the $m=0$ branch, for which\cite{Ezawa}
\begin{eqnarray}
{\bf f}_{0} &=& \sqrt{K \over 2 \pi \beta c {\sf A}} \bigg\{ \bigg[ i {\cal C} \, e^{- \gamma K z} + i \beta \, e^{-i \beta K z} \nonumber \\
&+& i \beta {\cal A} \, e^{i \beta K z} \bigg]  {\bf e}_{\rm K} + \bigg[ - \gamma {\cal C} \, e^{- \gamma K z} \nonumber \\
&+& i e^{-i \beta K z} - i {\cal A} \, e^{i \beta K z} \bigg] {\bf e}_z \bigg\} e^{i {\bf K} \cdot {\bf r}},
\label{0 modes}
\end{eqnarray}
where
\begin{equation}
\gamma(c) \equiv \sqrt{1 - (c/v_{\rm l})^2},
\end{equation}
\begin{equation}
{\cal A}(c) \equiv {(\beta^2 - 1)^2 - 4 i \beta \gamma \over (\beta^2 - 1)^2 + 4 i \beta \gamma},
\end{equation}
and
\begin{equation}
{\cal C}(c) \equiv { 4 \beta (\beta^2 - 1) \over (\beta^2 - 1)^2 + 4 i \beta
\gamma}.
\end{equation}
Note that $|{\cal A}(c)|=1$.

The $m = 0$ branch therefore contributes to Eq.~(\ref{substrate g definition}) the quantity
\begin{equation}
g_{ij}^{(0)}({\bf r},\omega) = {\omega^2 \delta_{ij} \over 8 \pi^2 v_{\rm t}^3}  \bigg[  (1-\delta_{iz}) \, I_{0}\big({\textstyle{\omega z \over v_{\rm t}}}\big) 
+ 2 \, \delta_{iz}  J_{0}\big({\textstyle{\omega z \over v_{\rm t}}}\big) \bigg],
\end{equation}
where
\begin{widetext}
\begin{eqnarray}
I_{0}(Z) &\equiv& \int_1^{1/\nu} \! {ds \over s^4 \sqrt{s^2-1}} \ \bigg| \bigg({4 (s^2-2) (s^2 - 1)^{1 \over 2} \over (s^2-2)^2 + 4 i (s^2 - 1)^{1 \over 2} 
(1-\nu^2 s^2)^{1 \over 2}}\bigg) e^{- Z \sqrt{1-\nu^2s^2} /s} \nonumber \\
&+& \big(s^2-1\big)^{1 \over 2} e^{-iZ \sqrt{s^2-1} /s} + \big(s^2-1\big)^{1 \over 2} \bigg({(s^2-2)^2 - 4 i (s^2-1)^{1 \over 2} (1-\nu^2s^2)^{1 \over 2} 
\over  (s^2-2)^2 + 4 i (s^2-1)^{1 \over 2} (1-\nu^2s^2)^{1 \over 2}}\bigg) e^{iZ \sqrt{s^2-1} /s} \bigg|^2
\label{I0 definition}
\end{eqnarray}
and
\begin{eqnarray}
J_{0}(Z) &\equiv& \int_1^{1/\nu} \! {ds \over s^4 \sqrt{s^2-1}} \ \bigg| \big(1-\nu^2 s^2\big)^{1 \over 2} \bigg({4 (s^2-2) (s^2 - 1)^{1 \over 2} \over (s^2-2)^2 + 4 i (s^2 - 1)^{1 \over 2} (1-\nu^2 s^2)^{1 \over 2}}\bigg) e^{- Z \sqrt{1-\nu^2s^2} /s} \nonumber \\
&-& i e^{-iZ \sqrt{s^2-1} /s} + i \bigg({(s^2-2)^2 - 4 i (s^2-1)^{1 \over 2} (1-\nu^2s^2)^{1 \over 2} \over  (s^2-2)^2 + 4 i (s^2-1)^{1 \over 2} 
(1-\nu^2s^2)^{1 \over 2}}\bigg) e^{iZ \sqrt{s^2-1} /s} \bigg|^2.
\label{J0 definition}
\end{eqnarray}
\end{widetext}
Furthermore, we note the asymptotic results
\begin{equation}
\lim_{Z \rightarrow \infty} I_{0}(Z) = {2 \over 3} \big(1-\nu^2\big)^{3/2}
\end{equation}
and
\begin{equation}
\lim_{Z \rightarrow \infty} J_{0}(Z) = {2 \over 3}\big(2+\nu^2\big)\sqrt{1-\nu^2}.
\end{equation}

\subsection{Rayleigh branch}

The normalized vibrational eigenmodes for the Rayleigh branch $m = {\rm R}$ are\cite{Ezawa} 
\begin{eqnarray}
{\bf f}_{\rm R} &=& \sqrt{K \over {\cal K} {\sf A}} \bigg\{ \bigg[i e^{-\varphi K z} - i \big({\textstyle{2 \varphi \eta \over 1 + 
\eta^2}}\big) \, e^{- \eta K z} \bigg] {\bf e}_{\rm K} \nonumber \\
&-&  \bigg[\varphi e^{- \varphi K z} - \big({\textstyle{2 \varphi \over 1 + \eta^2}} \big) \, e^{- \eta K z} \bigg] {\bf e}_z \bigg\} e^{i {\bf K} \cdot {\bf r}},
\label{R modes}
\end{eqnarray}
where
\begin{equation}
\varphi \equiv \sqrt{1 - (c_{\rm R}/v_{\rm l})^2} \ \ \ {\rm and} \ \ \ \eta \equiv \sqrt{1 - (c_{\rm R}/v_{\rm t})^2}.
\label{varphi and eta definitions}
\end{equation}
Furthermore,
\begin{equation}
{\cal K} \equiv {(\varphi - \eta)(\varphi - \eta + 2 \varphi \eta^2) \over 2 \varphi \eta^2}. 
\end{equation}

In Eq.~(\ref{varphi and eta definitions}), $c_{\rm R} $ is the velocity of the Rayleigh (surface) waves, given by 
\begin{equation}
c_{\rm R}= \xi \, v_{\rm t},
\end{equation}
where $\xi$ is the root between 0 and 1 of the polynomial
\begin{equation}
\xi^6 - 8 \xi^4 + 8(3-2\nu^2) \xi^2 - 16(1-\nu^2) = 0,
\label{Rayleigh velocity equation}
\end{equation}
with $\nu$ the velocity ratio defined above in Eq.~(\ref{nu definition}). In terms of $\xi$ we can write
\begin{equation}
\varphi = \sqrt{1 - \nu^2 \xi^2} \ \ \ {\rm and} \ \ \ \eta = \sqrt{1 - \xi^2}.
\label{varphi and eta simplifications}
\end{equation}

The Rayleigh branch contributes an amount
\begin{eqnarray}
g_{ij}^{({\rm R})}({\bf r},\omega) &=& {\omega^2 \delta_{ij} \over 4 \pi {\cal K} c_{\rm R}^3} \bigg[  (1-\delta_{iz}) \, 
I_{\rm R}\big({\textstyle{\omega z \over c_{\rm R}}}\big) \nonumber \\
&+&  2  \delta_{iz}  \, J_{\rm R}\big({\textstyle{\omega z \over c_{\rm R}}}\big) \bigg] ,
\end{eqnarray}
where
\begin{equation}
I_{\rm R}(Z) \equiv \big[ e^{-\varphi Z} - \big({\textstyle{2 \varphi \eta \over 1 + \eta^2}}\big) e^{- \eta Z} \big]^2 
\label{IR definition}
\end{equation}
and
\begin{equation}
J_{\rm R}(Z) \equiv \varphi^2 \big[  e^{-\varphi Z} - \big({\textstyle{2 \over 1 + \eta^2}}\big) e^{- \eta Z} \big]^2 .
\label{JR definition}
\end{equation}
Both $I_{\rm R}(Z)$ and $I_{\rm R}(Z)$ vanish in the large $Z$ limit.

\begin{widetext}

\subsection{Local DOS}

By symmetry, the local DOS tensor is diagonal, with the in-plane components $g_{xx}=g_{yy}$  generally different than the perpendicular component $g_{zz}$. Combining the results derived above, we obtain our principal result,
\begin{equation}
g_{ij}({\bf r},\omega) = \delta_{ij} {\omega^2 (2 \! + \! \nu^3) \over 6 \pi^2 v_{\rm t}^3} \bigg[ (1-\delta_{iz}) \ g_{\|}\! \big({\textstyle{\omega z \over v_{\rm l}}}\big) 
+  \delta_{iz} \ g_{\perp} \! \big({\textstyle{\omega z \over v_{\rm l}}}\big) \bigg],
\label{tensor decomposition}
\end{equation}
where
\begin{equation}
g_{\|}(\theta) \equiv  {3 \over 2+\nu^3} \, I_{\rm SH}\big(\theta/ \nu\big) + {3 \nu^3 \over 8(2+\nu^3)} \big[ I_{+}\big(\theta\big) \! + \! I_{-}\big(\theta\big) \big] 
+ {3 \over 4(2+\nu^3)} \, I_{0}\big(\theta/ \nu\big) + {3 \pi \over 2(2+\nu^3) {\cal K} \xi^3} \, I_{\rm R}\big(\theta/ \xi \nu\big)
\label{g par}
\end{equation}
and
\begin{equation}
g_{\perp}(\theta) \equiv {3 \nu^3 \over 4(2+\nu^3)} \big[ J_{+}\big(\theta\big) \! + \! J_{-}\big(\theta\big) \big] 
+ {3 \over 2(2+\nu^3)} \, J_{0}\big(\theta/ \nu\big) + {3 \pi \over (2+\nu^3) {\cal K} \xi^3} \, J_{\rm R}\big(\theta/ \xi \nu\big).
\label{g per}
\end{equation}
The prefactor $\omega^2 (2 \! + \! \nu^3) / 6 \pi^2 v_{\rm t}^3$ in Eq.~(\ref{tensor decomposition}) is the bulk Debye DOS. In the large $\theta$ limit, corresponding to high frequencies $\omega$, large perpendicular distances $z$, or both, the asymptotic results given above lead to
\begin{equation}
\lim_{\theta \rightarrow \infty} g_{\|}(\theta) = \lim_{\theta \rightarrow \infty} g_{\perp}(\theta) = 1,
\end{equation}
\end{widetext}
in which case the local DOS of Eq.~(\ref{tensor decomposition}) reduces to the Debye value given in Eq.~(\ref{bulk g}).

\section{application to silicon \label{Si section}}

\begin{figure}
\includegraphics[width=8.0cm]{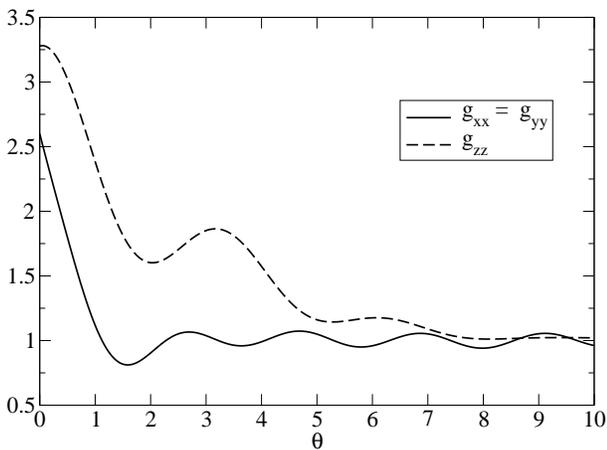}
\caption{\label{gtensor figure} Dimensionless local DOS functions $g_{\|}$ (solid curve) and $g_{\perp}$ (dashed curve), defined in Eqs.~(\ref{g par}) and (\ref{g per}), for an isotropic elastic continuum model of a Si substrate. Here $\theta \equiv \omega z / v_{\rm l}.$ The DOS functions are essentially equal to unity for $\theta > 10$.}
\end{figure}

To apply these results, we shall approximate Si as an isotropic elastic continuum with longitudinal and transverse sound velocities
\begin{eqnarray}
v_{\rm l} &=& 8.47 \! \times \! 10^5 \, {\rm cm \ s}^{-1}, \\
v_{\rm t} &=& 5.86 \! \times \! 10^5 \, {\rm cm \ s}^{-1},
\label{Si velocities}
\end{eqnarray}
and mass density
\begin{eqnarray}
\rho = 2.33  \, {\rm g \ cm}^{-3}.
\label{Si density}
\end{eqnarray}
The velocities above follow from using the measured elastic constants $c_{11} = 1.67 \! \times \! 10^{12} \ {\rm dyn \, cm^{-2}}$ and $c_{44} = 8.01 \! \times \! 10^{11} \ {\rm dyn \, cm^{-2}}$ at $70 \, {\rm K}$ reported in Ref.~[\onlinecite{Madelung}]. A more precise method of determining these velocities would be to use the actual anisotropic elastic tensor to construct a best fit to an isotropic one.\cite{Fedorov} However this only makes small changes in the final results.

The Rayleigh sound speed is determined from Eq.~(\ref{Rayleigh velocity equation}). The velocity ratio defined in Eq.~(\ref{nu definition}) is $\nu = 0.693,$ leading to $\xi = 0.882,$ and hence 
\begin{equation}
c_{\rm R} = 5.17 \! \times \! 10^5 \, {\rm cm \ s}^{-1}. \ \ \ \ ({\rm for \ Si})
\end{equation}
In addition, $\varphi = 0.792,$ $\eta = 0.472,$ and ${\cal K} = 0.610.$

In Fig.~\ref{gtensor figure}, we present plots of the dimensionless DOS functions $g_{\|}(\theta)$ and $g_{\perp}(\theta)$ for Si.  Here $\theta \equiv \omega z / v_{\rm l},$ where $\omega$ is the angular frequency, $z$ is the perpendicular distance from the surface, or depth, and $v_{\rm l}$ is the bulk longitudinal sound velocity. The dimensionless DOS has an interesting oscillatory structure when $\theta$ is small, and approaches unity in the large $\theta$ limit. When $\theta = 0$,
\begin{equation}
g_{\|}(0) = 2.60
\label{gpar(0)}
\end{equation}
and
\begin{equation}
g_{\perp}(0) = 3.28.
\label{gper(0)}
\end{equation}
On the substrate surface, $z \! = \! 0$, the local DOS is considerably modified from the bulk Debye value, as a consequence of Eqs.~(\ref{gpar(0)}) and (\ref{gper(0)}). In particular, $g_{xx}(0,\omega)$ and $g_{yy}(0,\omega)$ are quadratic in frequency with a constant of proportionality $g_{\|}(0)$ times that of the Debye model. Similarly, $g_{zz}(0,\omega)$ is quadratic in $\omega$, with a constant of proportionality a factor of $g_{\perp}(0)$ larger than in the Debye case.

\begin{figure}
\includegraphics[width=8.0cm]{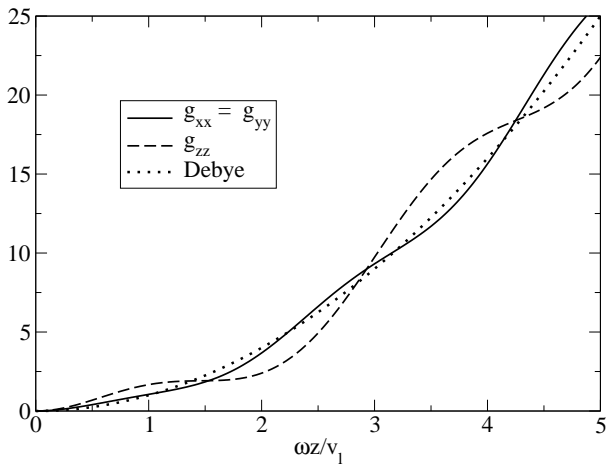}
\caption{\label{frequency figure} Frequency dependence of the local DOS $g_{xx}(z,\omega)=g_{yy}(z,\omega)$ (solid curve) and $g_{zz}(z,\omega)$ (dashed curve) for a fixed $z > 0$, compared to the purely quadratic frequency dependence (dotted curve) of the Debye model. In all cases the DOS is given in units of $(2 \! + \! \nu^3) / 6 \pi^2 v_{\rm t}^3$. The frequency is plotted in units of $v_{\rm l}/z$, where $v_{\rm l}$ is the bulk longitudinal sound velocity. Elastic parameters corresponding to Si are assumed.}
\end{figure}

The oscillatory structure in $g_{\|}(\theta)$ and $g_{\perp}(\theta)$ also modifies the frequency dependence of the local DOS inside the substrate. This is illustrated in Fig.~\ref{frequency figure}, where $g_{ij}(z,\omega)$ is plotted in units of $(2 \! + \! \nu^3) / 6 \pi^2 v_{\rm t}^3$. In these units, the Debye DOS is simply $\omega^2$, shown as the dotted line in the figure.

\section{discussion \label{discussion section}}

We have calculated the local, eigenfunction-weighted acoustic phonon DOS tensor for a model substrate consisting of a semi-infinite isotropic elastic continuum with a stress-free surface, and applied the results to Si. The local DOS on the surface is quadratic in $\omega$, with a proportionality constant that is enhanced compared to the three-dimensional bulk Debye model. The enhancement factors for Si are given in Eqs.~(\ref{gpar(0)}) and (\ref{gper(0)}). Inside the substrate the frequency dependence is non-quadratic, as illustrated in Fig.~\ref{frequency figure}. At depths $z$ large compared with $v_{\rm l}/\omega$, where $v_{\rm l}$ is the bulk longitudinal sound velocity, or at frequencies $\omega$ large compared with $v_{\rm l}/z$, the local DOS approaches the isotropic Debye value. These regimes corresponds to the large $\theta$ limits of $g_{\|}(\theta)$ and $g_{\perp}(\theta)$, defined in Eqs.~(\ref{g par}) and (\ref{g per}). 

As expected, the presence of the free surface significantly changes the vibrational properties of the system when the condition $z < \lambda$ is met, where $\lambda$ is a characteristic bulk phonon wavelength with frequency $\omega$. At a fixed frequency, this condition reduces to a requirement on distances from the surface, whereas at fixed depths it becomes a condition on frequency.

\begin{acknowledgments}

This work was supported by the National Science Foundation under CAREER Grant No.~DMR-0093217 and NIRT Grant No.~CMS-040403, and by the Research Corporation.

\end{acknowledgments}


\begin{thebibliography}{99}
\bibitem{definition footnote} As defined, this quantity is directly related to the imaginary part of the Fourier-transformed retarded phonon Green's function.
\bibitem{area footnote} Ezawa, in Ref.~[\onlinecite{Ezawa}], assumes an infinite lateral area ${\sf A}$. Therefore, we multiply his eigenfunctions by a factor $\sqrt{4 \pi^2/{\sf A}}$.
\end{thebibliography}
\end{document}